# BURST FLUENCE DISTRIBUTIONS OF SOFT GAMMA REPEATERS 1806-20 AND 1900+14 IN THE RXTE PCA ERA

ZACHARY PRIESKORN[1,2] AND PHILIP KAARET[1]
[1]Department of Physics and Astronomy, The University of Iowa, Iowa City IA 52242
[2]Department of Astronomy and Astrophysics, Pennsylvania State University, University Park PA 16802; **prieskorn@psu.edu**

## ABSTRACT

We study the fluence distributions of over 3040 bursts from SGR 1806-20 and over 1963 bursts from SGR 1900+14 using the complete set of observations available from the *Rossi X-Ray Timing Explorer*/Proportional Counter Array through March 2011. Cumulative event distributions are presented for both sources and are fitted with single and broken power laws as well as an exponential cutoff. The distributions are best fit by a broken power law with exponential cutoff, however the statistical significance of the cutoff is not high and the upper portion of the broken power law can be explained as the expected number of false bursts due to random noise fluctuations. Event distributions are also examined in high and low burst rate regimes and power law indices are found to be consistent, independent of the burst rate. The contribution function of the event fluence is calculated. This distribution shows that the energy released in the SGR bursts is dominated by the most powerful events for both sources. The power law nature of these distributions combined with the dominant energy dissipation of the system occurring in the large, less frequent bursts is indicative of a self-organized critical system (SOC), as suggested by Goğus, et al. in 1999.

*Keywords:* pulsars: individual (SGR 1806−20, SGR 1900+14) – stars: magnetars – stars: neutron – X-rays: bursts

## 1. INTRODUCTION

Soft gamma repeaters (SGRs) are associated with slowly rotating, extremely magnetized neutron stars (see reviews by Woods & Thompson **2006** and Mereghetti **2008**). They are sometimes found in young supernova remnants and are characterized by recurrent emission of short duration, ~0.1 s (Kouveliotou **1995**), bursts of soft gamma-rays/hard x-rays with spectra described by optically thin thermal bremsstrahlung (OTTB) at kT ~ 20-40 keV (Goğus et al. **2001**). Recent studies at lower energies, <15 keV, have found that an OTTB model is no longer accurately fitted to the spectra and there is a rollover which can be best fit with two black bodies or alternatively a cutoff powerlaw (Feroci et al. **2004**; Olive et al. **2004**; Nakagawa et al. **2007**; Israel et al. **2008**; Lin et al. **2011**; van der Horst et al. **2012**). There is evidence that these bursts are the result of starquakes that occur when the neutron star (NS) crust fractures due to a build up of stress associated with the evolution of the strong magnetic field, greater than $10^{14}$ G (Thompson & Duncan **1995**).

In Thompson & Duncan's seminal 1995 paper first describing magnetars they give a theoretical description of the repeating bursts observed from SGRs. The ultra-strong magnetic fields rotating with the core of the NS are pulled through the crust, causing stress to build up. When the stresses become more than the crust can support it fractures. An electron/positron pair plasma is ejected from the star and captured by the magnetic field near the NS surface. This plasma then cools via emission of soft gamma-rays and X-rays.

The statistical properties of SGR bursts have been shown to be similar to those observed for earthquakes (Cheng et al. **1996**; Gutenberg & Richter **1956**, **1965**). Event energy distributions from SGR 1806-20 and SGR 1900+14 sources have been well fit with a power law described by $dN \propto E^{-\gamma} dE$ and an exponent, γ, of 1.6 - 1.7 (Goğus et al. **1999**, **2000**). Many natural systems have been found to fit power laws (Aschwanden **2011**). This power law distribution of energy release was recognized to occur in many natural systems that exhibit non-linear energy dissipation and is commonly referred to as a self-organized critical (SOC) system (Bak et al. **1987**).

Here, we report on burst energy distributions using all of the data available from the *Rossi X-ray Timing Explorer* (*RXTE*) for two of the most active SGRs, SGR 1806-20 and SGR 1900+14, covering the period from 12/1996 – 02/2011. The initial search found 3040 bursts from SGR 1806-20 and 1963 from SGR 1900+14. Differential and cumulative event fluence distributions are presented for each source. These were fit with single power laws as reported by past authors (Laros et al. **1987**; Woods et al. **1999**; Goğus et al. **1999**) but we also fitted single and broken power laws with exponential cutoffs and were able to improve upon the quality of fit for each source. We also use the large number of bursts observed to define periods of high and low burst activity. The burst fluence distribution for high and low burst phases were then analyzed. The contribution function was calculated for the burst fluence distributions. This function would be peaked at a characteristic fluence if one existed in the data set. This distribution shows the contribution of the different fluence bins to the total energy released by the bursts. In §**2** of this paper the observations for each source are described, §**3** covers analysis and the results, and §**4** presents our conclusions.

## 2. TARGETS AND OBSERVATIONS

SGR 1806-20 has been observed to exhibit sporadic bursting behavior since 1983 (Laros et al. **1987**). BATSE (launched April 1991) detected more sporadic bursting behavior. It entered an active phase in 1996. Two weeks of observations with *RXTE/*PCA led to the discovery of 7.47 s pulsations and confirmed its nature as a magnetar (Kouveliotou et al. 1998). This source remained relatively inactive until it released a giant flare in 2004 (Hurley et al. 2005). The luminosity of this flare had a peak of a few times $10^{47}$ erg s$^{-1}$ (Hurley et al. 2005; Palmer et al. 2005). The source remained active, emitting hundreds of smaller bursts for the next 2 years, although it generally decreased in activity over that time.

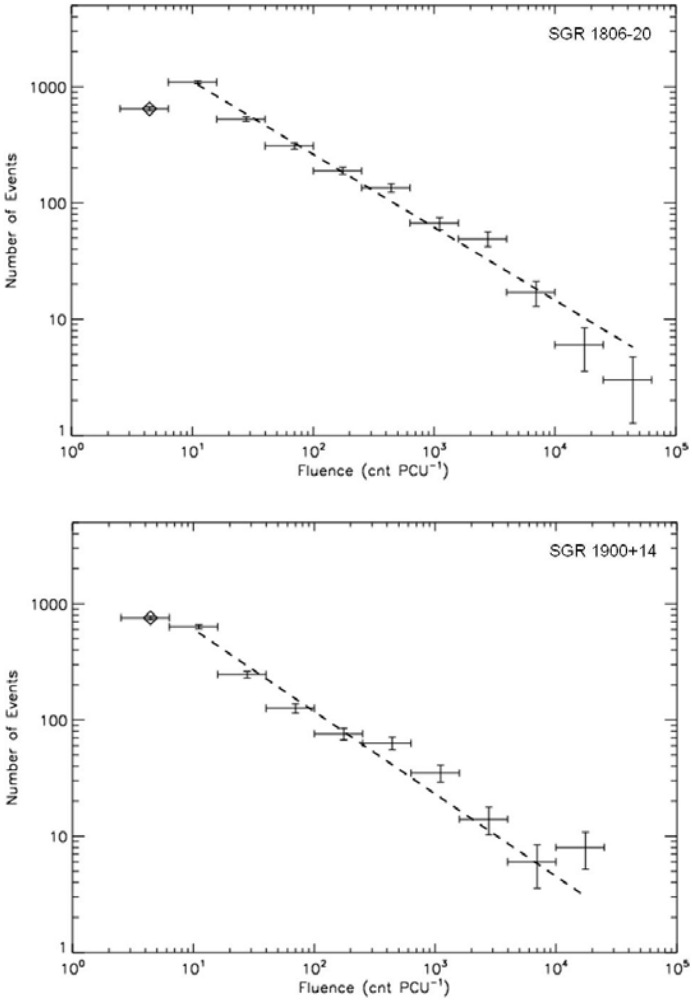

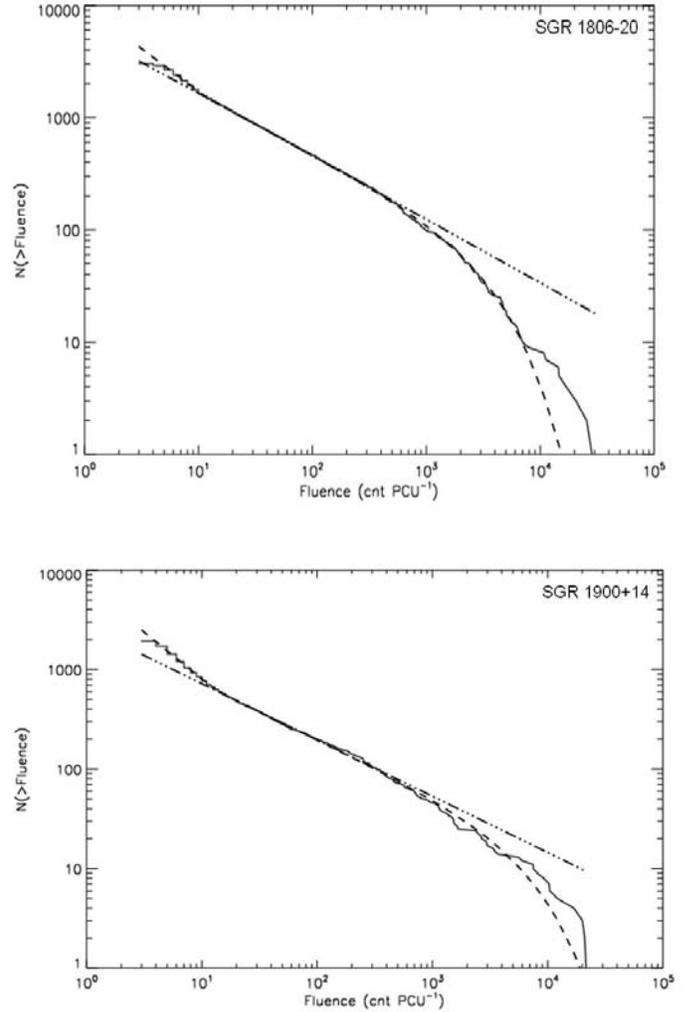

**Figure 1.** Differential burst fluence distributions for SGR 1806-20 (top) and SGR 1900+14 (bottom). Both are well fitted by a power law of the form $dN \propto E^{-\gamma} dE$, with $\gamma = 0.63\pm0.04$ for SGR 1806-20 and $1.71\pm0.06$ for SGR 1900+14. The lowest fluence bin was excluded from the fit due to decreased detector sensitivity and increased sensitivity to background fluctuations.

**Figure 2**. The solid line is data, dash-dot is a single power law $N(>E) \propto E^{1-\gamma}$ and the dashed is a broken power law plus exponential cutoff, $E^{1-\gamma} e^{\alpha x}$. The exponential cutoff is, however, not significant and the high fluence fall-off is likely from a poor sampling of the upper fluence events. The excess at low fluence, fit by the upper power law, corresponds to the number of expected false detections of bursts during the 3 Ms of observations for each source, §3.2. The fits are consistent with a single power law over 3 orders of magnitude, a fingerprint of self-organized critical (SOC) behavior.

SGR 1900+14 became active in May 1998 after a period of long inactivity (Kouveliotou et al. 1993). It then emitted a giant flare in August of 1998 (Hurley et al. 1999). The persistent emission from this source shows pulsations with a period of 5.2 s, matching the period of pulsations observed in the tail of the 1998 giant flare (Cline, Mazets & Golenetskii 1998; Hurley 1999; Feroci 2001). The source has remained active over the last decade but no further giant flares have been observed.

The PCA onboard *RXTE* observed SGR 1806-20 and SGR 1900+14 for a total of more than 800 hours, or 3 Ms, each. An automated burst search, similar to the searches by Woods et al. (1999) and Goğus et al. (1999) was performed for all of the available data. We used standard-1 (2-60 keV) data constrained to observations 5° above the Earth's horizon. Each 0.125 s bin was searched. A background count rate was estimated using 5 s of data 3 s before and after the bin being searched. Bins in excess of 125 counts bin$^{-1}$ were assumed to contain burst emission and were excluded from the background estimate. A burst was defined as any continuous set of bins in which the count rate exceeded 5.5 σ the average local background rate. The count fluence was then estimated by integrating the background-subtracted counts for all bins included in the burst. Bursts which happen very close in time were counted as a single burst if their fluence did not drop below the 5.5 σ of the average local background rate and counted as multiple bursts when they did. There were a small number of bursts which saturated the *RXTE*/PCA and were left out of our study because the burst fluence cannot be accurately calculated for these events. We report the fluence in PCA counts PCU$^{-1}$ because no spectral analysis was done in this work. A conversion factor, 2 x $10^{-12}$ ergs cm$^{-2}$ counts$^{-1}$, can be used to determine the burst fluence in ergs cm$^{-2}$. This was reported by Goğus et al. (1999, 2000) based on BATSE spectral measurements.

**Table 1.**
Fit parameters and quality of fit for cumulative SGR burst fluence distributions.

| | Fit Parameters | | | | | |
|---|---|---|---|---|---|---|
| **SGR 1806-20** | $\gamma_1$ | $\gamma_2$ | *Break* (counts PCU$^{-1}$) | $\alpha$ | KS | KS Probability |
| *All Bursts 3040 bursts* | | | | | | |
| Single power law | - | 1.58±0.03 | - | - | 0.086 | 0.000 |
| Broken power law | 1.77±0.01 | 1.54±0.01 | 12 | - | 0.007 | 0.999 |
| Broken power law with exponential cutoff | 1.77±0.01 | 1.54±0.01 | 12 | $-2.2 \pm 0.47 \times 10^{-4}$ | 0.007 | 0.999 |
| *High Burst Rate 2232 bursts* | | | | | | |
| Single power law | - | 1.59±0.03 | - | - | 0.070 | 0.000 |
| Broken power law | 1.75±0.01 | 1.56±0.01 | 12 | - | 0.008 | 0.999 |
| Broken power law with exponential cutoff | 1.75±0.01 | 1.56±0.01 | 12 | $-2.4 \pm 0.59 \times 10^{-4}$ | 0.008 | 0.999 |
| *Low Burst Rate 323 bursts* | | | | | | |
| Single power law | - | 1.59±0.03 | - | - | 0.188 | 0.000 |
| Broken power law | 2.42±0.04 | 1.59±0.02 | 8 | - | 0.013 | 0.999 |
| Broken power law with exponential cutoff | 2.42±0.04 | 1.59±0.02 | 8 | $-1.0 \pm 0.41 \times 10^{-4}$ | 0.013 | 0.999 |
| **SGR 1900+14** | | | | | | |
| *All Bursts 1963 bursts* | | | | | | |
| Single power law | - | 1.64±0.04 | - | - | 0.152 | 0.000 |
| Broken power law | 1.94±0.03 | 1.56±0.02 | 14 | - | 0.014 | 0.945 |
| Broken power law with exponential cutoff | 1.94±0.03 | 1.56±0.02 | 14 | $-1.2 \pm 0.10 \times 10^{-4}$ | 0.014 | 0.945 |
| *High Burst Rate 1626 bursts* | | | | | | |
| Single power law | - | 1.63±0.04 | - | - | 0.126 | 0.000 |
| Broken power law | 1.85±0.03 | 1.55±0.02 | 14 | - | 0.015 | 0.971 |
| Broken power law with exponential cutoff | 1.85±0.03 | 1.55±0.02 | 14 | $-1.7 \pm 0.11 \times 10^{-4}$ | 0.016 | 0.951 |
| *Low Burst Rate 236 bursts* | | | | | | |
| Single power law | - | 1.73±0.04 | - | - | 0.273 | 0.000 |
| Broken power law | 2.48±0.04 | 1.57±0.02 | 10 | - | 0.019 | 0.999 |
| Broken power law with exponential cutoff | 2.48±0.04 | 1.57±0.02 | 10 | $-8.5 \pm 1.6 \times 10^{-4}$ | 0.019 | 0.999 |

**Notes.** Multiple models were fit for each source distribution including a single power law ($N(>E) \propto E^{1-\gamma_2}$), broken power law ($E < $ *Break*: $N(>E) \propto E^{1-\gamma_1}$ and $E \geq $ *Break*: $N(>E) \propto E^{1-\gamma_2}$) and broken power law with exponential cutoff, where the power law above the break is multiplied by $exp(\alpha E)$.

## 3. ANALYSIS AND DISCUSSION

We fitted the data for each source using differential and cumulative distributions. The data was initially fitted with a single power law but for both types of distribution the fit parameters indicated that the likelihood of the model representing the data were very low, reduced $\chi^2$ values between 3 – 5 and a *Kolmogorov-Smirnov* (KS) probability of zero for both sources. Examining the fluence distributions of each source (Figure 1 and Figure 2), it is apparent that at the lower and upper ends of the distribution there are more components to the fit. We then fit a broken power law and exponential cutoff using the cumulative distributions. The cumulative distribution retains all of the original information from the data set and more clearly shows the excess and fall off from the power law. The method and results for each source, distribution and different fits are described in the following section.

### 3.1. Differential Distributions of Burst Fluence

The events from each data set were binned in equally spaced logarithmic fluence steps ($dN/d$ log$E$). The number of bins for each distribution were determined by using the Sturges formula (Sturges 1926), given by $log_2N+1$ where $N$ is the number of events in the distribution. Using a least-squares fitting method, the data are well fit by a power law distribution over 3 orders of magnitude in fluence. The power law exponent, $\gamma$, was found to be 0.63 ± 0.04 for SGR 1806-20 and 0.71 ± 0.06 for SGR 1900+14 where $dN \propto E^{-\gamma} dE$, see Figure 1. The exponent depends strongly on where the lower threshold was set for thresholds near the minimum detectable fluence but as mentioned above, none of the single power law fits show evidence of being good fits to the data with reduced $\chi^2$ values between 3 – 5. We excluded the first point in the distribution, shown as a diamond, as this represents events only a few counts above the threshold and thus in the limited sensitivity range of the detector.

The differential distribution for SGR 1806-20 shows a drop off below the single power law in the highest fluence bins. We attempted to fit a broken power law and single power law with exponential cutoff to account for this feature but neither provides a significantly improved fit due to the low statistics at the high fluence end of the distribution. The inability to fit this feature as well as the lowest data bin included in the fit motivated using the cumulative distribution described in §3.2. The values for a single power law fit to the differential distribution are consistent with those reported by Goğus et al. (1999, 2000).

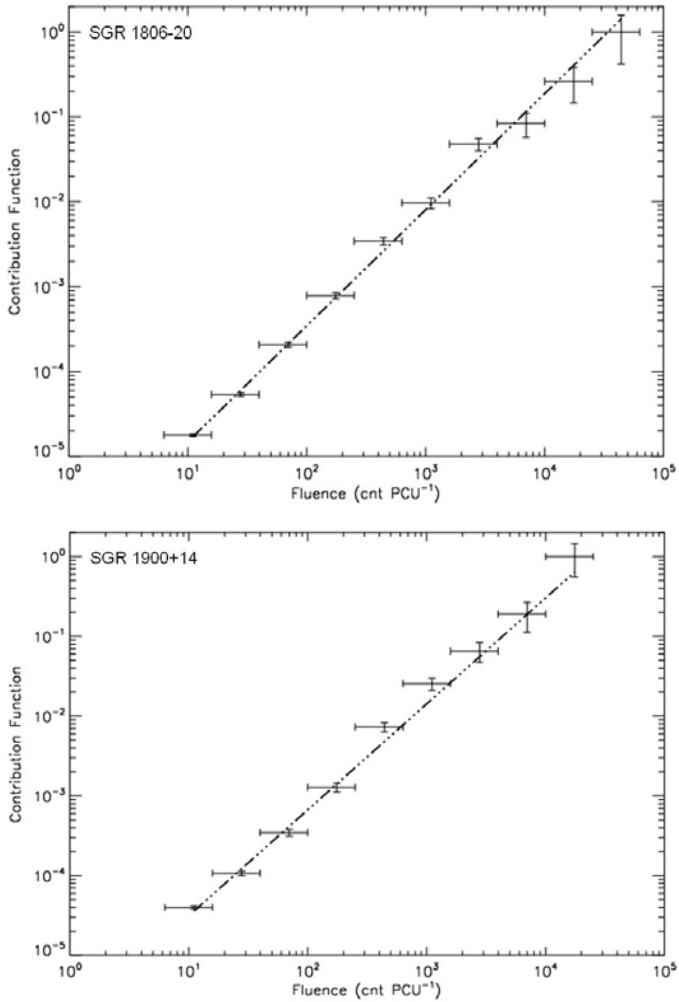

**Figure 3.** The fluence distribution was binned and the contribution function, $(dN/d\log E)\,E$, was calculated. The distribution is well fitted by a single power law over 4 orders of magnitude and shows that while the highest fluence bursts occur least often, they dominate the energy release in the system, another sign of SOC behavior. The normalized contribution is shown.

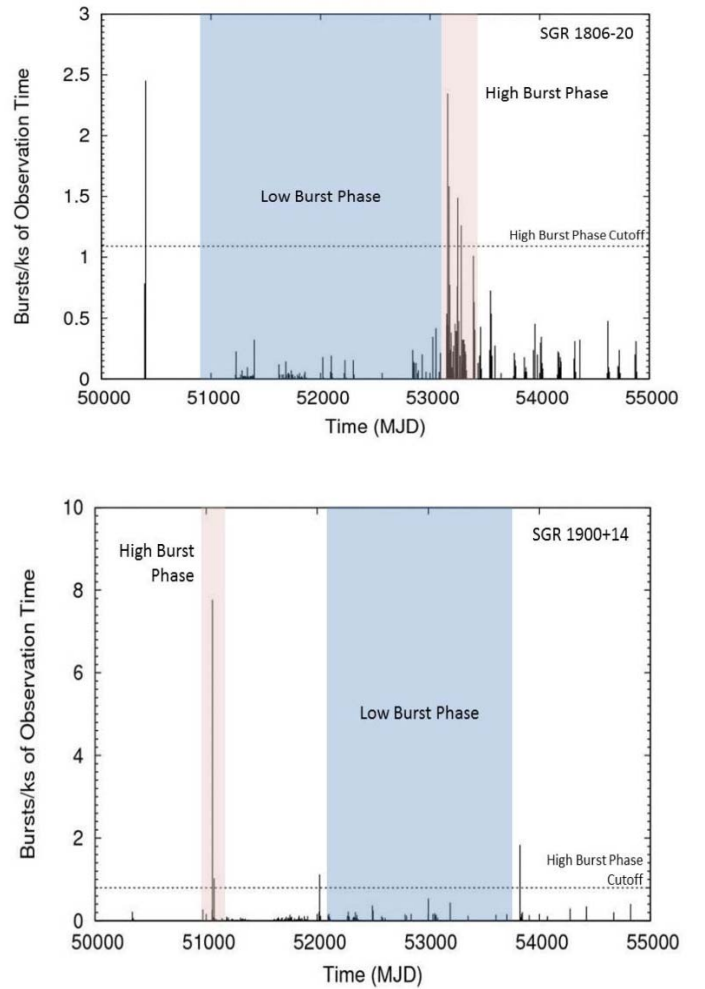

**Figure 4.** The number of bursts per ks of observation over 3 Ms of observations for SGR 1806-20 (top) and SGR 1900+14 (bottom). The high burst activity is defined as any observation where the burst rate exceeds that average rate for all of the observations, approximately 1 burst/kilosecond for each source. Fluence distributions were made for each phase, shown in the shaded regions, and are presented in Figure 5.

### 3.2. Cumulative Distributions of Burst Fluence

Cumulative distributions were calculated and fitted with a single power law using the maximum likelihood method. The uncertainty in the cumulative distribution was determined from the Poisson uncertainty in the binned data, resulting in using the square root of the sum of all the counts greater than a given fluence. The differential power law described in §3.1 is related to that fitted to the cumulative distribution by the following, $N(>E) \propto E^{1-\gamma}$. The single power law fit to the cumulative distribution above a low fluence cutoff of 5 counts PCU$^{-1}$, as described above being due to low detector sensitivity, have KS probabilities of zero. Examining the cumulative distribution, Figure 2, the large excess at low fluence and drop off at high fluence clearly stand out. Fitting a broken power law for both sources dramatically improves the quality of fit, KS probabilities of 0.999 for SGR 1806-20 and 0.945 for SGR 1900+14. For SGR 1806-20 $\gamma = 1.54 \pm 0.01$ and for SGR 1900+14 $\gamma = 1.56 \pm 0.02$. Adding an exponential cutoff to account for the high fluence drop off, $N(>E) \propto E^{1-\gamma}\exp(\alpha E)$, failed to improve the quality of the fit significantly. Each fit, with and without the exponential cutoff, has the same KS probability out to 5 significant figures. Figure 2 shows multiple fits to the cumulative distribution for each source and Table 1 gives fit parameters and goodness of fit results for each source and model.

The need for a lower fluence power law in addition to the power law covering the middle 3 orders of magnitude in the distribution is due to the large excess above a single power law fit to the other 3 orders of magnitude. The power law at lower fluence has a $\gamma = 1.77 \pm 0.01$ for SGR 1806-20 and $\gamma = 1.94 \pm 0.03$ for SGR 1900+14. Assuming the detector noise fits a Poisson distribution we estimate that for 3 Ms of observing time with 0.125 s bins there should be ~360 false bursts detected due to noise fluctuations above the 5 counts PCU$^{-1}$ cutoff. This is on the order of the excess in bursts observed for each source which have different numbers of bursts but almost the same amount of observing time. Fitting a high burst rate and low burst rate distribution, as described in §3.4, gives more support to this explanation for the low fluence excess. The low burst rate distribution, which contains very few bursts but the majority of the observing time, is seen to have most of the excess events, Figure 5.

The falloff at high fluence which appears visually in the cumulative distribution of each source, although more exaggerated for SGR 1806-20, can be fit with the inclusion of an exponential cutoff with $\alpha = -2.2 \pm 0.47 \times 10^{-4}$ for SGR 1806-20 and $\alpha = -1.2 \pm 0.10 \times 10^{-4}$ for SGR 1900+14. However, the goodness of fit remains unchanged because the KS value is dominated by the largest difference between model and data, which occurs at a lower fluence where the power law fit dominates[1]. Goğus et al. (**1999**, **2000**) exclude the high fluence events from their fits with the argument that the intrinsic distribution is under sampled. The visual fit with the exponential component appears better than the fit without, suggesting that the drop off in high fluence events is real, however the low statistics do not allow this to be determined definitively. If the drop off is related to a characteristic fluence at approximately 30,000 count PCU$^{-1}$, it may be difficult to confirm until there are significantly more events due to the low statistics at higher fluence.

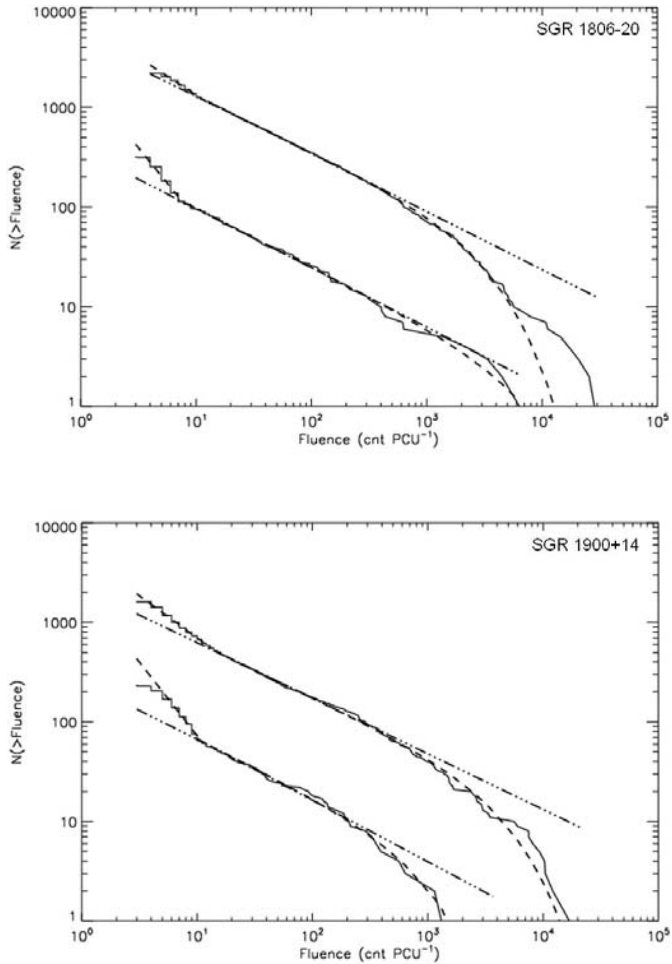

**Figure 5.** Cumulative distributions for low (bottom curve) and high (upper curve) rates of burst activity fit with single power law, $N(>E) \propto E^{1-\gamma}$, and broken power law with exponential cutoff, $E^{1-\gamma} e^{\alpha x}$. The fits are consistent with those calculated for the entire distribution. The event fluence is thus independent of the rate of bursts, an important fingerprint of self-organized critical (SOC) behavior. Table **1** gives fit parameters and the number of bursts included for each source and model.

### 3.3. Contribution Function Distribution

To better understand the nature of these events and how the energy release is distributed, we examined the contribution function, (d$N$/d $logE$) $E$, for the fluence distributions. Using the differential distribution we calculated the contribution function and plotted it against d $logE$. This distribution should peak at the characteristic fluence of the intrinsic distribution, see Figure **3**. Instead we observe a distribution well fit by a single power law over 4 orders of magnitude, with exponents of 1.37 ± 0.03 and 1.34 ± 0.05 for SGR 1806-20 and SGR 1900+14 respectively. This further supports the idea that the falloff observed in the cumulative fluence distribution is not related to the characteristic fluence of the sources. Both distributions indicate that the largest portion of energy release occurs in the highest fluence bursts instead of the more numerous low fluence bursts.

---
[1] The KS value is defined as the maximum value of the absolute difference between two cumulative distributions and appears as follows: $KS = \max_{-\infty < x < \infty} |S_N(x) - P(x)|$ (Press et al. **1989**).

### 3.4. Event Distributions as a Function of Burst Activity

The large number of detected bursts allows us to distinguish between different levels of burst activity for each source. Figure **4** shows the burst rate for all observations of each source We define high burst activity as a burst rate exceeding the average rate for all observations, approximately 1 burst per ks of observing time for each source. Cumulative fluence distributions were made for separate periods of burst activity for each source and fit with a single power law and a broken power law with exponential cutoff as described in §**3.2**. The burst rate dependent cumulative distributions are shown in Figure **5**. The exponents for each source are the following: SGR 1806-20 high phase is 1.56 ± 0.02 and low phase is 1.59 ± 0.03; SGR 1900+14 high phase is 1.55 ± 0.02 and low phase is 1.57 ± 0.02. The fit parameters and goodness of fit values are summarized in Table **1**. The power law exponent for each phase is consistent with that of the total bursts indicating that the burst fluence statistics are independent of the level of burst activity and over the time for which we have observations with *RXTE*/PCA, approximately 15 years. The fact that the excess of events at low fluence is dominant in the low burst rate regime is supportive of the idea that this excess is due to random fluctuations in the background, as suggested in §**3.2**.

## 4. CONCLUSIONS

The collected *RXTE*/PCA observations for SGR 1806-20 and SGR 1900+14 provide thousands of bursts for each source to perform statistical analysis on the fluence distributions. We fitted the cumulative distributions for SGR 1806-20 and SGR 1900+14 with a single power law, broken power law and both with exponential cutoffs. The distributions were best fit with a broken power law with exponential cutoff but there is not a significant statistical difference between that and the broken power law without a cutoff. Fitting all the features of the distribution allowed us to improve the uncertainty on the fit parameters for a single power law over 3 orders of magnitude, giving $\gamma$'s of 1.54 ± 0.01 and 1.56 ± 0.02 for SGR 1806-20 and SGR 1900+14 respectively.

The contribution function was calculated for each source and was found to be well fit by a single power law over 4 orders of magnitudes of fluence. The function is clearly not peaked, thus we do not observe a characteristic fluence in these distributions. It also shows how energy release in repeating bursts is dominated by the emission from high fluence events instead of the more numerous low fluence events.

The burst activity for each source varies significantly over the ~3 Ms of observations. We found the power law fit for data at high and low burst rate and find that the exponent for different levels of activity are consistent with the exponent for the total event distribution. This indicates that the fluence of individual bursts is independent of burst activity, however the rate of bursts can vary significantly over time.

The power law fits to the event fluence distributions when all the events are included as well as the rate dependent distributions, especially when considered over 3 orders of magnitude, provide strong evidence that the repeating bursts from SGRs are the result of a self-organized critical (SOC) system. The independence of the power law fit from burst rate is another indicator of SOC behavior. A SOC system requires a constant source of energy, in SGRs this can be attributed to the ultra-strong magnetic fields. The undoubtedly complex nature of the interactions occurring in the crust of a NS combined with this constant source of power in the magnetic field then explain why we observe SOC behavior from SGRs.

We thank Ken Gayley for his suggestion to calculate the contribution function and for his help in doing so. We would also like to thank Joanne Hill, Randall McEntaffer and Craig Kletzing for their comments. Fotis Gavriil also offered important advice for this work.

## REFERENCES

Aschwanden, M. 2011, Self-Organized Criticality in Astrophysics. (Heidelberg, Germany: Springer)
Bak, P., Tang, C., Wiesenfeld, K. 1987, PRL, 59, **381**
Cheng, B., Epstein, R., Guyer, R., et al. 1995, Nature, 382, **518**
Cline, T., Mazets, E., & Golenetskii, S. V. 1998, IAU Circ., 7002, 1
Feroci, M. G., Caliandro, E., Massaro, S., et al. 2004. ApJ, 612, **408**
Feroci, M. G., Hurley, K., Duncan, R. C., & Thompson, C., 2001, ApJ, 549, **1021**
Goğus, E., Kouveliotou, C., Woods, P. M., et al. 2001, ApJ, 558, **228**
Goğus, E., Woods, P. M., Kouveliotou, C., et al. 1999, ApJ, 526, **L93**
Goğus, E., Woods, P. M., Kouveliotou, C., et al. 2000, ApJ, 532, **L121**
Gutenberg, B., & Richter, C. F. 1956, Bull. Seism. Soc. Am., 46, 105
Gutenberg, B., & Richter, C. F. 1965, Seismicity of the Earth and Associated Phenomena (New York: Hafner)
Hurley, K., Boggs, S. E., Smith, D. M., et al. 2005, Nature, 434, **1098**
Hurley, K., Cline, T., Mazets, E., et al. 1999, Nature, 397, **41**
Israel, G L, Romano, P., Mangano, V., et al. 2008, ApJ, 685, **1114**
Kouveliotou, C. 1995, Ap&SS, 231, **49**
Kouveliotou, C., Dieters, S., Strohmayer, T., et al. 1998, Nature, 393, **235**
Kouveliotou, C., Fishman, G. J., Meegan, C. A., et al. 1993, Nature, 362, **728**
Laros, J. G., Fenimore, E. E., Klebesadel, R. W., et al. 1987, ApJ, 320, **L111**
Lin, L., Kouveliotou, C., Goğus, E., et al. 2011, ApJ, 739, **87**
Mereghetti, S. 2008, Astron Astrophys Rev, 15, **225**
Nakagawa, Y. E., Yoshida, A., Hurley, K., et al., 2007. PASJ, 59, **653**
Olive, J, Hurley, K., Sakamoto, T., et al. 2004, ApJ, 616, **1148**
Palmer, D.M. et al. 2005, Nature, 434, **1107**
Press, W., Flannery, B., Teukolsky, S., et al. 1989, Numerical Recipes. Fortran Ed. (Cambridge: Cambridge University Press)
Sturges, H. A. 1926, J. ASA, 65
Thompson, C., & Duncan, R. C. 1995, MNRAS, 275, **255**
van der Horst, A. J., Kouveliotou, C., Gorgone, N. M., et al. 2012, ApJ, 749, **122**
Woods, P., Kouveliotou, C., van Paradijs, J., et al. 1999, ApJ, 524, **L55**
Woods, P., Thompson, C. 2006, in Compact Stellar X-ray Sources, ed. W. Lewin & M. van der Klis (Cambridge, UK: Cambridge University Press), **547**